# Title

Comprehensive interrogation of gene lists from genome-scale cancer screens with oncoEnrichR

# Authors

Sigve Nakken[1,2,3,*], Sveinung Gundersen[3], Fabian L. M. Bernal[4], Dimitris Polychronopoulos[5], Eivind Hovig[1,3] and Jørgen Wesche[1,2]

# Affiliations

[1] Department of Tumor Biology, Institute for Cancer Research, Oslo University Hospital, Norway

[2] Centre for Cancer Cell Reprogramming, Institute of Clinical Medicine, Faculty of Medicine, University of Oslo, Norway

[3] Centre for Bioinformatics, Department of Informatics, University of Oslo, Norway

[4] University Center for Information Technology, University of Oslo, Norway

[5] Early Data Science and Computational Oncology, Research and Early Development, Oncology R&D, AstraZeneca, Cambridge, UK.

*To whom correspondence should be addressed:

Email: sigven@ifi.uio.no, Tel: +4795753022

*Abstract*

Genome-scale screening experiments in cancer produce long lists of candidate genes that require extensive interpretation for biological insight and prioritization for follow-up studies. Interrogation of gene lists frequently represents a significant and time-consuming undertaking, in which experimental biologists typically combine results from a variety of bioinformatics resources in an attempt to portray and understand cancer relevance. As a means to simplify and strengthen the support for this endeavor, we have developed oncoEnrichR, a flexible bioinformatics tool that allows cancer researchers to comprehensively interrogate a given gene list along multiple facets of cancer relevance. oncoEnrichR differs from general gene set analysis frameworks through the integration of an extensive set of prior knowledge specifically relevant for cancer, including ranked gene-tumor type associations, literature-supported proto-oncogene and tumor suppressor gene annotations, target druggability data, regulatory interactions, synthetic lethality predictions, as well as prognostic associations, gene aberrations, and co-expression patterns across tumor types. The software produces a structured and user-friendly analysis report as its main output, where versions of all underlying data resources are explicitly logged, the latter being a critical component for reproducible science. We demonstrate the usefulness of oncoEnrichR through interrogation of two candidate lists from proteomic and CRISPR screens. oncoEnrichR is freely available as a web-based workflow hosted by the Galaxy platform (https://oncotools.elixir.no), and can also be accessed as a stand-alone R package (https://github.com/sigven/oncoEnrichR).

*Significance:* oncoEnrichR is a comprehensive and user-friendly report engine for the interpretation and prioritization of candidate hits generated by genome-scale screens in cancer.

## 1 Introduction

The search for novel cancer-implicated genes is currently fueled by data-driven high-throughput genome-scale screening studies, exemplified by the various types of gene perturbation experiments possible with the CRISPR/Cas9 technology (1). Most high-throughput screens typically produce the same conceptual output, in the form of dozens or hundreds of candidate genes ranked by effect sizes. A significant challenge in this respect is to aggregate sufficient prior knowledge regarding the candidate hits, both individually and at a systems level, in order to enable an optimal prioritization of genes for in-depth experimental validation. General gene set enrichment resources are frequently used for this task, as well as portals that permit exploration of large-scale cancer omics data (2–5). Nevertheless, users often end up with a fragmented approach, where they need to combine results from a diverse set of annotation tools (5). The whole interpretation process is generally impeded by a lack of dedicated frameworks that can systematically prioritize a collection of candidate hits with respect to their cancer relevance, a key element in the interrogation of hits from screens conducted in a cancer setting.

We have developed *oncoEnrichR*, a user-friendly gene set interpretation workflow in which a diverse suite of annotation and prioritization modules highlight the cancer relevance of candidate hits, both quantitatively through global and tumor-type specific rank scores, but also qualitatively through multiple annotations that underscore potential tumorigenic roles and drug-target opportunities. The gene-centric modules of oncoEnrichR are supplemented with system-level features that offer insights from a candidate list at the level of signaling pathways, molecular gene signatures, as well as regulatory interactions and the protein interactome. A structured and interactive gene set report, highly configurable by the user, is readily available as the main output per analysis.

Here we describe the various analysis modules available in oncoEnrichR, and demonstrate through use cases how it can function as a valuable platform for candidate hit interpretation in cancer.

## 2  Materials and Methods

*Data collection*

The backend of oncoEnrichR is based on the integration of cancer-relevant properties of human genes and their interrelationships from > 20 publicly available knowledge resources. The main objective here is to provide a breadth of large-scale molecular datasets that cover key dimensions in our current comprehension of cancer relevance (a complete overview is provided in **Supplementary Table 1**). Conceptually, the datasets can be broadly categorized as *i) interactions*, such as protein-protein or transcription factor-target interactions, *ii) collections*, like protein complexes and biological pathways, and *iii)* basic *annotations*, including descriptive types, for instance tumor suppressor gene or cancer hallmark annotation, and quantitative types, such as those derived from large-scale omics datasets, e.g. percent tumor samples in TCGA with a homozygous deletion for a given gene. The preparation of all data sources is outlined in detail in **Supplementary Methods.**

*Gene-cancer association rank*

We exploited data from the powerful Open Targets Platform (OTP) to create a quantitative ranking of genes with respect to cancer relevance. OTP provides an aggregated association score (range 0-1) between phenotypes (in the form of Experimental Factor Ontology (EFO) terms) and human genes based on a range of evidence sources, e.g. genetic associations, text mining, or data from animal models (6,7). In order to identify association scores restricted to cancer phenotypes, we developed a semi-manually curated mapping between EFO terms and 32 primary tumor types/tissues (breast, colon/rectum, lung etc.), using OncoTree as the starting point (8). This mapping allowed us further to compute a scaled rank with respect to the association between any given gene and a primary tumor type, and also globally, i.e. pan-cancer (see **Supplementary Methods,** and **Supplementary Figure 1)**

*Data Availability Statement*

All source code for oncoEnrichR is publicly available at https://github.com/sigven/oncoEnrichR. User documentation is provided at https://sigven.github.io/oncoEnrichR. The raw molecular datasets that are integrated in the data backend of oncoEnrichR, as well as code for preprocessing procedures, are available at https://doi.org/10.5281/zenodo.7116316

## 3  Results

*Architecture*

The oncoEnrichR gene set interpretation tool has been developed as an R package, and functions conceptually as a workflow of four main steps (**Figure 1**). The web interface to oncoEnrichR is facilitated by the Galaxy platform (https://oncotools.elixir.no), which ensures a reproducible and collaborative analysis framework (9).

*Input and output*

The main user input to the tool is a basic list of human gene identifiers, with an upper limit of 500 entries. To cater for ease of use, multiple commonly used identifier types are permitted, e.g. official gene symbols, UniProt accessions, as well as transcript, gene, or protein identifiers (RefSeq/Ensembl). Importantly, oncoEnrichR can also map historical gene aliases towards their up-to-date primary symbol designations, a feature that is often necessary in the processing of user-generated gene lists.

An optional specification of a background gene set for use in gene set enrichment analysis is further available, which ensures that enrichment results can be interpreted correctly with respect to the nature of the underlying screen. The user can also configure multiple parameters of the individual analysis modules, exemplified by significance thresholds in enrichment analysis or confidence thresholds for queried protein-protein interactions. Finally, the user can flexibly choose any combination of modules to be run and included as output in the final report, in that sense allowing the user to control the focus and scope of the gene set analysis.

The output of the oncoEnrichR workflow is two-fold. A stand-alone, structured HTML report is provided as the main output, enabling user interaction with the resulting visualizations and ranked tables provided for each individual analysis module (outlined below). Moreover, a multi-sheet Excel workbook with all annotations, interactions and enrichment results is readily available. To cater to transparency, all versions of underlying software and databases are provided in the output files.

*Analysis modules*

A candidate gene set can currently be interrogated with oncoEnrichR through 16 different analysis modules (**Supplementary Table 1**). In general, each module considers the candidate set through a particular perspective on gene function or cancer relevance, and makes up an individual section in the output report. A small, educational user guide is populated for each section in the report, giving the user a basic introduction to the type of analysis performed and underlying databases used. Here, we briefly outline the various types of modules that are offered by oncoEnrichR, with a focus on what they provide regarding functional insights and cancer relevance for the gene set in question.

**Gene function, tumor-type associations, and drug-target opportunities**

A significant number of genes in the human genome are still poorly characterized with respect to function, and members of this set typically show up in candidate lists from high-throughput screens. Given that such uncharacterized genes provide a level of novelty for potential follow-up experiments, oncoEnrichR features a dedicated *Poorly characterized genes* module that highlights and ranks such entries, essentially considering genes with a lack of curated Gene Ontology (GO) annotations or gene summary descriptions.

In the *Cancer associations* module, the user can interrogate the candidate hits with respect to current knowledge on genes with tumorigenic roles. Genes are classified as tumor suppressor genes, proto-oncogenes, or potential cancer drivers, utilizing support from multiple resources (**Supplementary Methods**, 10–13). All genes are ranked within the candidate set according to their overall strength of association to cancer, and where the association metric is based upon aggregated evidence from multiple data types, such as text mining, genetic associations, or animal models (**Figure 2A**). A dedicated heatmap is furthermore showing the relative strength of association of candidate genes towards distinct tumor types (**Figure 2B**). A separate *Cancer hallmarks* module indicates whether the different hallmarks of cancer are promoted or suppressed by members of the candidate set, including links to associated literature. Targeted drugs, both in early and late clinical development, that are specifically indicated for one or more cancer conditions, are listed in a *Drug association* module. This module also allows the user to prioritize the complete set of candidates with respect to their drug targeting potential, based on comprehensive target tractability data.

The workflow contains two additional modules that can aid the identification of therapeutic targets within the candidate set. In the *Gene fitness effects* module, the tool shows which candidate targets are required for cellular fitness in different molecular contexts, lending upon data from large-scale CRISPR–Cas9 whole-genome dropout screens in human cancer cell lines (14). Importantly, the intersection of candidate hits with such genetic screen data also indicates relevant cell lines to be used in experimental follow-up studies. The integration of fitness effects with target tractability data and known molecular biomarkers allows for further prioritization of candidate hits as therapeutic targets (**Figure 2C**). On top of that, oncoEnrichR incorporates recent machine learning predictions on synthetic lethality in cancer cell lines, effectively allowing the prioritization of candidate hits that are involved in such interactions, either as internal interactions within the candidate set or through interactions with other genes (15).

**Functional enrichment, subcellular localization, and molecular interactions**

Exploring the candidate set through a systems perspective can often give important biological insights, and has become a common approach for the interpretation of gene lists, particularly within complex diseases such as cancer. In oncoEnrichR, system level analyses of the candidate set are provided through a series of modules. A *Functional enrichment* module offers enrichment/over-representation analysis of the candidate gene set towards Gene Ontology (GO), multiple pathway databases, and established cancer-relevant gene signatures (**Supplementary Table 2**). As mentioned above, the user can supply a dedicated background gene set to be used in the enrichment analysis, and configure enrichment significance thresholds.

An interactive protein-protein interaction (PPI) network of the candidate set is provided based on available data from the STRING resource (16). Here, the tool shows both internal interactions within the candidate set, and the most significant interactions with non-candidate set proteins. Importantly, the user can configure the confidence threshold of retrieved interactions. In the network provided by the *PPI* module, the roles of cancer genes are explicitly color-coded, and the user may also opt to attach targeted cancer drugs to the nodes in the network (**Figure 2D**). The PPI module further allows local community structures in the network to be interrogated, and all proteins in the candidate set are ranked according to their level of centrality (i.e. number of interactions) in the network.

A *protein complex* module further considers both curated and predicted protein complexes from multiple databases, catering for a ranking of the most cancer-relevant protein complexes with respect to candidate set members, including links to supporting literature. A *subcellular compartment* module offers annotation of the subcellular localization patterns for proteins in the query set, in which a subcellular anatogram provides an effective heatmap that indicates the most common subcellular compartments among proteins found in the candidate list (**Figure 2E**).

Considering the importance of inter- and intracellular signaling in cancer, oncoEnrichR integrates data on curated ligand-receptor interactions, as well as data on genes involved in transcriptional regulation. The latter set of interactions can be shown through a network, indicating the presence of directed transcription-factor (TF)-target relationships among members of the candidate set (**Figure 2F**).

**Gene mutation frequencies, co-expression patterns, and prognostic associations**

Through the use of large-scale genomics data from The Cancer Genome Atlas (TCGA), candidate hits can be interrogated and prioritized for aberration frequencies in different tumor types (somatic copy number events and point mutations/indels), a well-known indication of cancer relevance (17). The user can further examine the presence of known somatic mutation hotspots within candidate genes, and identify the amino acid sites that harbor loss-of-function variants in tumor samples. Genes in the candidate set that are found significantly co-expressed with known cancer genes in various tumor types are further highlighted in a separate section.

Expression profiling data from healthy tissues (Human Protein Atlas (HPA) and Genotype-Tissue Expression Project) are included in oncoEnrichR to portray tissue- and cell-type specific expression patterns for the candidate genes, which in turn may reveal enrichment of particular tissues/cell-types (18,19).

It is well known that expression levels and genomic aberrations of particular protein-coding genes are associated with overall patient survival in a number of tumor types (20). In the *Prognostic associations* module, oncoEnrichR shows both favorable and unfavorable prognostic associations of hits in the candidate set, considering either expression or methylation levels, or mutation or copy-number status of these genes across tumor samples.

*Use cases*

To showcase the usability of oncoEnrichR in the biological interpretation of gene lists, we explored the output of the tool using two candidate hit lists coming from CRISPR and protein proximity screens, respectively (complete output reports are available at https://doi.org/10.5281/zenodo.7115732).

**CRISPR screen: resistance to EGFR inhibition**

We initially interrogated a list of n = 57 hits originating from a CRISPR/Cas9 screen looking for novel drug resistance genes in non-small cell lung cancer (21). Of note, the set of 57 hits came here out of a domain-specific recommendation system, subsequently assessed by five independent domain experts with regard to their relevance as potential drivers of resistance to EGFR inhibition. The aim of the oncoEnrichR analysis in this context was thus to showcase how the tool can validate the relevance of known resistance markers, and add supporting evidence and perspectives on previously unknown markers (see also Figure 3 in (21)).

The report produced with oncoEnrichR confirms the strong cancer relevance of well-established resistance markers, including multiple known tumor suppressor genes (e.g. TP53, PTEN, NF1/2, SMARCA4) and proto-oncogenes (e.g. KRAS, ERBB2, MET, MAPK1, CDK4), as can be seen in the *Cancer associations* section. Three of the previously unknown resistance markers (CIC, EZH2 and CREBBP), are indicated to carry both oncogenic and tumor suppressive roles, a matter which can be further explored in the linked supporting literature from CancerMine, and also through entries shown in the *Cancer Hallmarks* evidence section. The tumor type-specific association overview (**Figure 3A**) provides further an opportunity to find candidates for which the current available evidence of association to lung cancer is weak (e.g. LZTR1 and FOSL1), and also other candidates which are strongly associated to selected tumor types (e.g. CYP1A1 in prostate cancer, SQSTM1 in sarcoma).

The *Drug associations* section highlights the availability of a large collection of anti-cancer drugs, which confirms how several of the unknown resistance markers can be targeted by various inhibitors (e.g. Tazemetostat (EZH2), Infigratinib/Erdafitinib (FGFR4)). An important consideration with respect to potential drug targets is their level of interactivity with other proteins, and through its interactome hub score calculation oncoEnrichR highlights the centrality of the SRC protein (**Figure 3B**). Furthermore, oncoEnrichR allows the interrogation of predicted synthetic lethality of multiple candidate hits, here exemplified through MAPK1/MAPK3 and CREBBP/EP300 interactions (**Figure**

3C). Finally, exploring potential prognostic associations can provide important supplementary evidence for selection of follow-up candidates. oncoEnrichR indicates that high expression of three candidate hits are significantly associated with poor survival in lung cancer (**Figure 3D**), interestingly also FOSL1, for which there was weak overall evidence in the *Cancer Associations* section.

**Protein proximity screen: FGFR1 interaction network**

We have previously performed a proximity labeling BioID experiment using FGFR1 (fibroblast growth factor receptor 1) as the bait, where the potential interaction partners were analyzed by Enrichr and BioGrid at the time (22). As aberrant FGF signaling is implicated in multiple tumor types (23,24), we wanted to use oncoEnrichR to reassess the dataset and further elucidate FGFR1 networks in cancer.

Proteins associating with FGFR1 upon stimulation with the activating ligand (FGF1) were subjected to oncoEnrichR analysis (n=74 proteins). oncoEnrichR successfully imported gene names from the Maxquant proteomic output file. As the screen was done several years ago, a few names were now obsolete (n=6), but they were all successfully updated by oncoEnrichR with the current official gene names (*Query verification* module).

Several features in oncoEnrichR proved useful for the validation of screen hits. For example, the *Protein-protein interaction* module revealed previously identified interactions. In the case of FGFR1, the proximal interaction network included the previously reported interactions of FRS2 and PLCG1 (**Figure 4A**). Moreover, FGFR1 is known to localize to the secretory pathway, the plasma membrane, the endosomal system and the nucleus. These localizations were clearly reflected in the subcellular anatogram showing that the majority of the hits in the BioID screen localized to these compartments (**Figure 4B**). As such, the subcellular compartment module may help in validating the results from the screen, but also point to new cell localizations for baits, via their interaction partners. For example, the reported *focal adhesion sites* could imply a new localization and function of FGFR1 with respect to focal adhesions. Finally, gene set enrichment analyses indicated the expected protein networks (e.g. secretory pathway, endosomes, clathrin-coated vesicle), as well as some new unexplored functions (e.g. focal adhesions).

A useful feature in oncoEnrichR is the annotation of hits as tumor suppressors or oncogenes, including previously generated knowledge from different tumor models. oncoEnrichR also informs on druggability and suggests matching inhibitors which can be used to validate and further investigate hits. For instance, oncoEnrichR reports on the cancer associations of IDO1 (indoleamine 2,3-dioxygenase 1), a strong hit in the screen, and suggest inhibitors that can be used to target this protein (e.g. Epacadostat and Linrodostat), providing an easy way to test its role in biological experiments (**Figure 4C**). Interestingly, CCNE1, an important oncogene involved in cell cycle progression, could possibly also be involved in FGFR1 signal transduction, as it was found as a strong hit in the screen. Large-scale CRISPR/Cas9 screen data found in the *Gene fitness effects* module can be used to suggest which cancer cell lines are dependent on your genes of interest. For instance, in the case of CCNE1, cell lines derived from ovary cancer depend on this gene, and oncoEnrichR shows a list of cell lines that can be used for further studies. Of note, CCNE1 is amplified in ovary cancer (21% of cases), as shown in the *Tumour aberration frequencies* module (**Figure 4D**).

oncoEnrichR also lists proteins where no or little evidence for function have been found, thereby highlighting proteins that could be the starting point for additional research projects. In the FGFR1 dataset, we identified several proteins with a poorly defined function (e.g. CSTPP1, C2CD4C, PRR14L and CRACDL, **Figure 4E**). Interestingly, C2CD4C was found to be frequently deleted in several cancers (e.g. 6% in cervix cancer), which may suggest a potential role as a tumor suppressor.

## 4     *Discussion*

Building upon existing data integration frameworks and multiple large-scale omics datasets, we have developed a feature-rich gene set interpretation tool that allows researchers to systematically interpret and prioritize long lists of candidate genes for cancer relevance. The frequently time-consuming nature of gene list interpretation can in part be attributed to a scarcity of single tools that comprehensively organize existing knowledge in the context of cancer. In this regard, we believe that oncoEnrichR, through its broad harvest of prior molecular knowledge and its multifaceted and interactive analysis report, fills an important gap. Moreover, the dual availability of the tool (i.e. web and command-line), makes it attractive for use both by researchers with a non-computational background, and for bioinformaticians that are developing complete screening analysis pipelines.

The output of a gene set analysis performed with oncoEnrichR reflects the current, temporary state of molecular cancer knowledge harvested from publicly available databases, for which update frequencies vary considerably (see **Supplementary Table 2**). Considering future maintenance and updates of oncoEnrichR, we have established and made available a suite of data preprocessing and quality control procedures that can be run on a regular basis (see *Data Availability Statement*). Furthermore, the modular nature of the tool and output report sections will simplify the potential addition of new functionality that can further enhance the mapping of cancer relevance.

Like any gene set interpretation tool, we acknowledge that the user needs to interpret the output carefully with reference to the screen that produced the candidate hits. While the simple, unranked gene list input makes oncoEnrichR easy to use, we recognize that an input list that incorporates effect size or rank from the originating screen will open the way for more sophisticated approaches of hit prioritization.

Importantly, genome-scale screens in cancer have many different objectives, and while oncoEnrichR is not tailored to a specific type of screen or designed to rank the hits according to a particular scientific question, we believe that the multiple perspectives provided by the tool can offer significant interpretation and prioritization support across a wide range of screening applications.

*Author contributions*

Design and concept: SN, JW. Data preparation and software implementation: SN. Galaxy platform infrastructure: FB, SG, EH. Testing and use case analysis: JW, DP. Wrote the manuscript: SN. All authors read, commented on, and approved the final manuscript.

*Acknowledgements*

*Funding:* This work was supported by the Research Council of Norway through its Centers of Excellence funding scheme [grant number: 262652]. JW was funded by the Norwegian Cancer Society (project number 198093). The authors wish to acknowledge the University Center for Information Technology (USIT) at the University of Oslo for infrastructure support, ELIXIR Norway for support with Galaxy development, Carlos Company (AstraZeneca UK) for feedback on the manuscript, and Dr. Ultan McDermott (Oncology R&D, AstraZeneca UK) for valuable feedback and input on functionality.

*Competing interests*

DP is a full-time employee and shareholder of AstraZeneca.

*References*

1. Bock C, Datlinger P, Chardon F, Coelho MA, Dong MB, Lawson KA, et al. High-content CRISPR screening. Nature Reviews Methods Primers. Nature Publishing Group; 2022;2:1–23.

2. Cerami E, Gao J, Dogrusoz U, Gross BE, Sumer SO, Aksoy BA, et al. The cBio cancer genomics portal: an open platform for exploring multidimensional cancer genomics data. Cancer Discov. 2012;2:401–4.

3. Tang Z, Kang B, Li C, Chen T, Zhang Z. GEPIA2: an enhanced web server for large-scale expression profiling and interactive analysis. Nucleic Acids Res. 2019;47:W556–60.

4. Huang DW, Sherman BT, Lempicki RA. Systematic and integrative analysis of large gene lists using DAVID bioinformatics resources. Nat Protoc. 2009;4:44–57.

5. Zhou Y, Zhou B, Pache L, Chang M, Khodabakhshi AH, Tanaseichuk O, et al. Metascape provides a biologist-oriented resource for the analysis of systems-level datasets. Nat Commun. 2019;10:1523.

6. Malone J, Holloway E, Adamusiak T, Kapushesky M, Zheng J, Kolesnikov N, et al. Modeling sample variables with an Experimental Factor Ontology. Bioinformatics. 2010;26:1112–8.

7. Ochoa D, Hercules A, Carmona M, Suveges D, Gonzalez-Uriarte A, Malangone C, et al. Open Targets Platform: supporting systematic drug-target identification and prioritisation. Nucleic Acids Res. 2021;49:D1302–10.

8. Kundra R, Zhang H, Sheridan R, Sirintrapun SJ, Wang A, Ochoa A, et al. OncoTree: A Cancer Classification System for Precision Oncology. JCO Clin Cancer Inform. 2021;5:221–30.

9. Afgan E, Baker D, Batut B, van den Beek M, Bouvier D, Cech M, et al. The Galaxy platform for

accessible, reproducible and collaborative biomedical analyses: 2018 update. Nucleic Acids Res. 2018;46:W537–44.

10. Sondka Z, Bamford S, Cole CG, Ward SA, Dunham I, Forbes SA. The COSMIC Cancer Gene Census: describing genetic dysfunction across all human cancers. Nat Rev Cancer. 2018;18:696–705.

11. Repana D, Nulsen J, Dressler L, Bortolomeazzi M, Venkata SK, Tourna A, et al. The Network of Cancer Genes (NCG): a comprehensive catalogue of known and candidate cancer genes from cancer sequencing screens. Genome Biol. 2019;20:1.

12. Lever J, Zhao EY, Grewal J, Jones MR, Jones SJM. CancerMine: a literature-mined resource for drivers, oncogenes and tumor suppressors in cancer. Nat Methods. 2019;16:505–7.

13. Martínez-Jiménez F, Muiños F, Sentís I, Deu-Pons J, Reyes-Salazar I, Arnedo-Pac C, et al. A compendium of mutational cancer driver genes. Nat Rev Cancer. Nature Publishing Group; 2020;20:555–72.

14. Dwane L, Behan FM, Gonçalves E, Lightfoot H, Yang W, van der Meer D, et al. Project Score database: a resource for investigating cancer cell dependencies and prioritizing therapeutic targets. Nucleic Acids Res. Oxford Academic; 2020;49:D1365–72.

15. De Kegel B, Quinn N, Thompson NA, Adams DJ, Ryan CJ. Comprehensive prediction of robust synthetic lethality between paralog pairs in cancer cell lines. Cell Syst. 2021;12:1144–59.e6.

16. Von Mering C, Jensen LJ, Snel B, Hooper SD, Krupp M, Foglierini M, et al. STRING: known and predicted protein--protein associations, integrated and transferred across organisms. Nucleic Acids Res. Oxford University Press; 2005;33:D433–7.

17. The Cancer Genome Atlas Research Network, Weinstein JN, Collisson EA, Mills GB, Mills Shaw KR, Ozenberger BA, et al. The Cancer Genome Atlas Pan-Cancer analysis project. Nat Genet. Nature Publishing Group; 2013;45:1113–20.

18. Jain A, Tuteja G. TissueEnrich: Tissue-specific gene enrichment analysis. Bioinformatics. 2019;35:1966–7.

19. Uhlen M, Oksvold P, Fagerberg L, Lundberg E, Jonasson K, Forsberg M, et al. Towards a knowledge-based Human Protein Atlas. Nat Biotechnol. 2010;28:1248–50.

20. Smith JC, Sheltzer JM. Genome-wide identification and analysis of prognostic features in human cancers. Cell Rep. 2022;38:110569.

21. Gogleva A, Polychronopoulos D, Pfeifer M, Poroshin V, Ughetto M, Martin MJ, et al. Knowledge graph-based recommendation framework identifies drivers of resistance in EGFR mutant non-small cell lung cancer. Nat Commun. 2022;13:1667.

22. Kostas M, Haugsten EM, Zhen Y, Sørensen V, Szybowska P, Fiorito E, et al. Protein Tyrosine Phosphatase Receptor Type G (PTPRG) Controls Fibroblast Growth Factor Receptor (FGFR) 1 Activity and Influences Sensitivity to FGFR Kinase Inhibitors. Mol Cell Proteomics. 2018;17:850–70.

23. Ahmad I, Iwata T, Leung HY. Mechanisms of FGFR-mediated carcinogenesis. Biochim Biophys Acta. 2012;1823:850–60.

24. Tanner Y, Grose RP. Dysregulated FGF signalling in neoplastic disorders. Semin Cell Dev Biol. 2016;53:126–35.

25. Dingar D, Kalkat M, Chan P-K, Srikumar T, Bailey SD, Tu WB, et al. BioID identifies novel c-MYC interacting partners in cultured cells and xenograft tumors. J Proteomics. 2015;118:95–111.

*Figures*

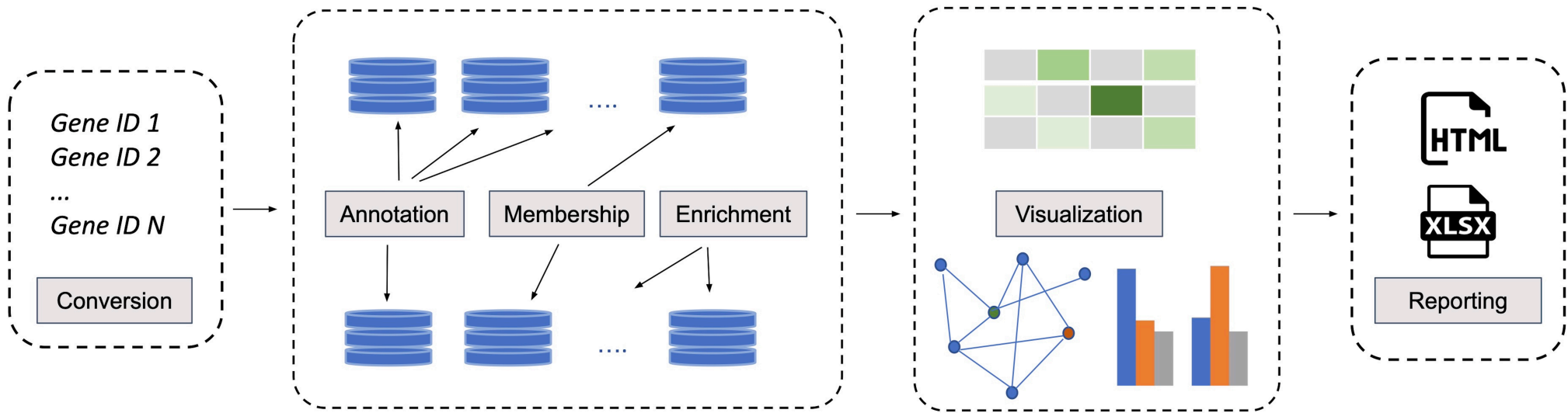

**Figure 1: Schematic overview of the oncoEnrichR workflow.** The tool consists of four key processing steps: 1) conversion and harmonization of gene/protein query identifiers, 2) annotation, membership and enrichment analyses of the query set against a comprehensive collection of cancer-relevant databases, 3) visualization of analysis results, and 4) report generation, either as HTML or Excel.

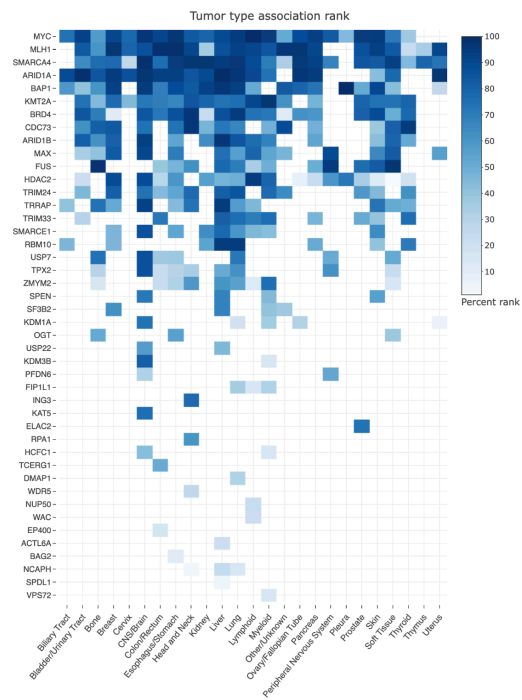
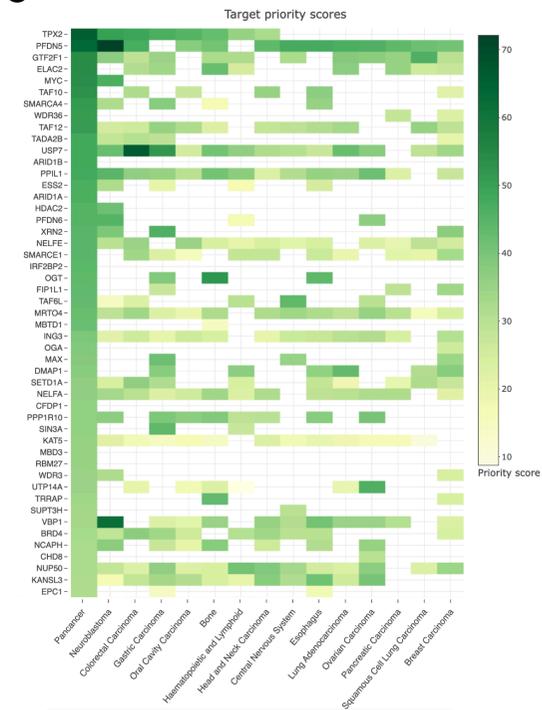
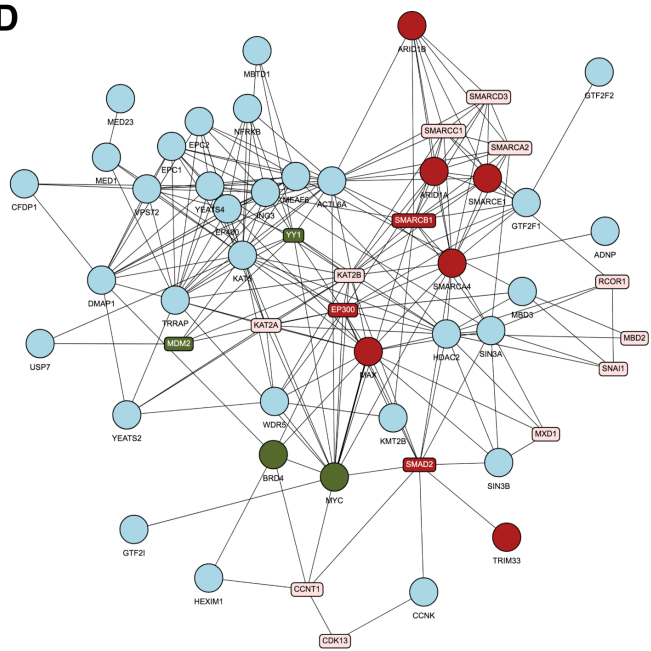
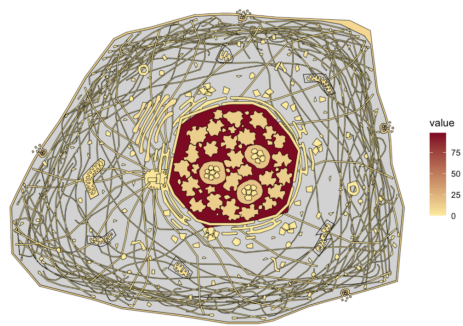
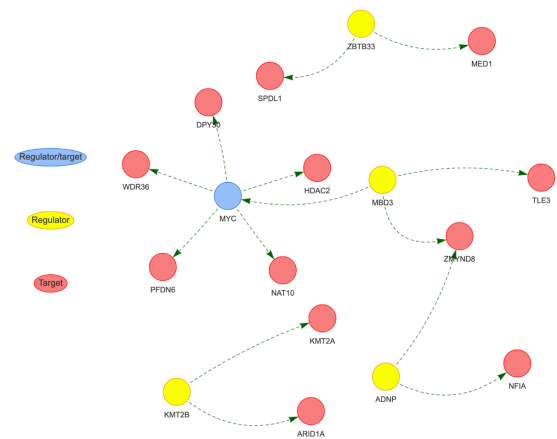

**Figure 2**: **Examples of analysis output from the oncoEnrichR workflow**. Output views from oncoEnrichR analysis modules when analyzing a set of n = 134 c-MYC interacting partners (25). The complete oncoEnrichR report for this example gene set is available for download at https://doi.org/10.5281/zenodo.7115745. **A**. Overall rank of query set proteins with respect to cancer relevance, considering the built-in pan-cancer gene rank score of oncoEnrichR. Users can also interrogate/filter the query set with respect to classifications as proto-oncogene/tumor suppressor genes, and view supporting literature. **B**. A heatmap of cancer relevance among query set members, measured on a per tumor-type basis according to oncoEnrichR's rank score. **C**. Target priority scores of query set members from Project Score, considering the combination of gene fitness scores in cancer cell lines and the availability of known biomarkers. **D**. An interactive protein-protein interaction network of query set members (shaped as circles), expanded with the most important interacting non-query set proteins (squares). Cancer gene roles are color-coded (tumor suppressors in green, proto-oncogenes in red, dual roles in black), and targeted drugs can be attached to the network. **E**. A subcellular heatmap of query set members, showcasing the most abundant subcellular compartments for the hits analyzed. **F**. Network of known regulatory interactions (transcription factor-target) for members of the query set.

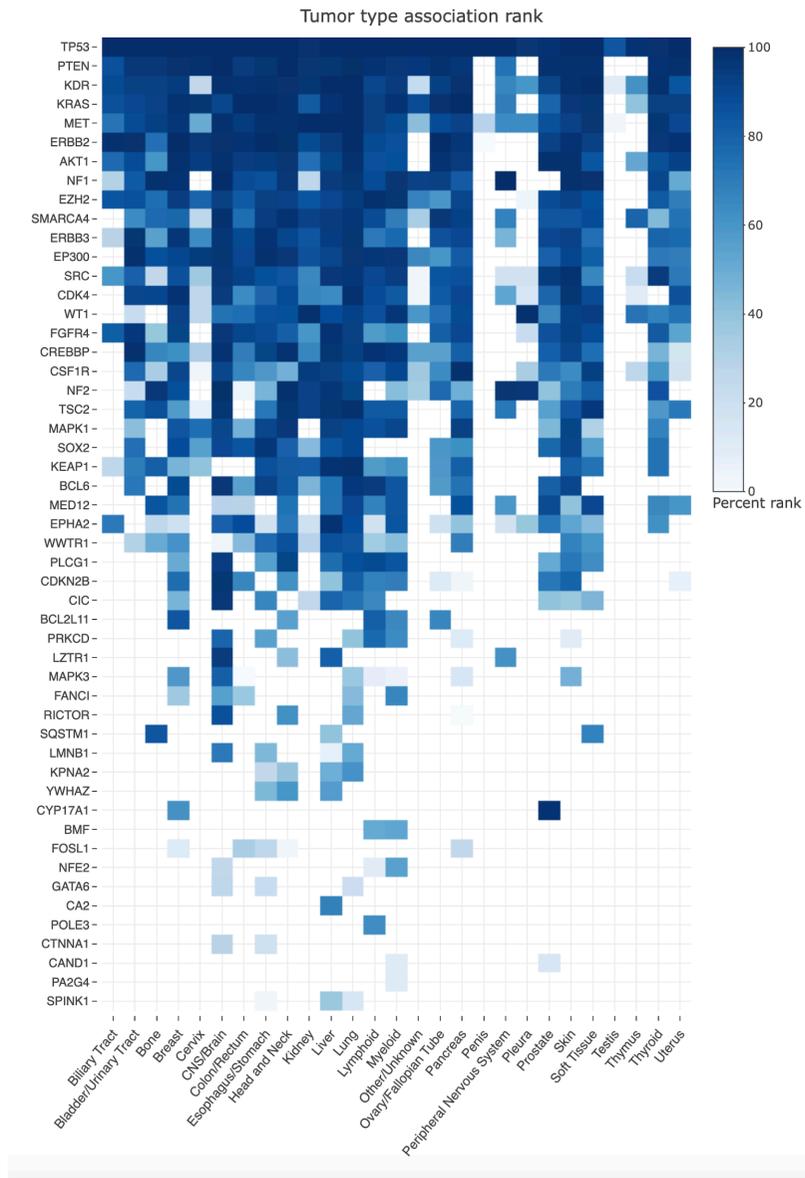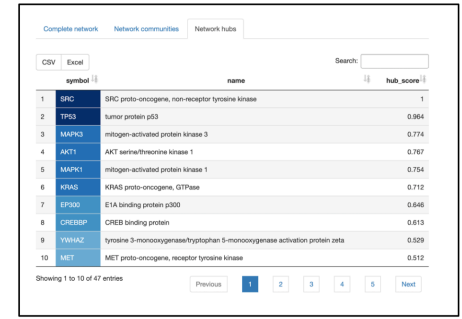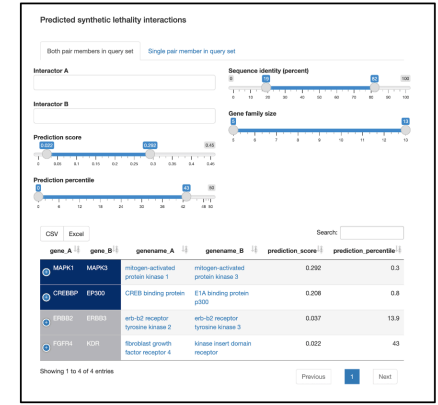

**Figure 3. CRISPR use case analysis - EGFRi resistance drivers.** Specific output views from the oncoEnrichR HTML report, analyzing N = 57 candidate drivers of EGFRi resistance (complete reports available at https://doi.org/10.5281/zenodo.7115732). **A.** Tumor-type association rank for all candidate hits found, confirming the strong cancer relevance of multiple known resistance markers. **B.** Ranking of candidate hits according to protein-protein network centrality scores, indicating the vast interactome of SRC. **C.** Predicted synthetic lethality interactions among members of the candidate set. **D.** Prognostic gene expression associations for candidate genes in lung cancer patients.

**A**

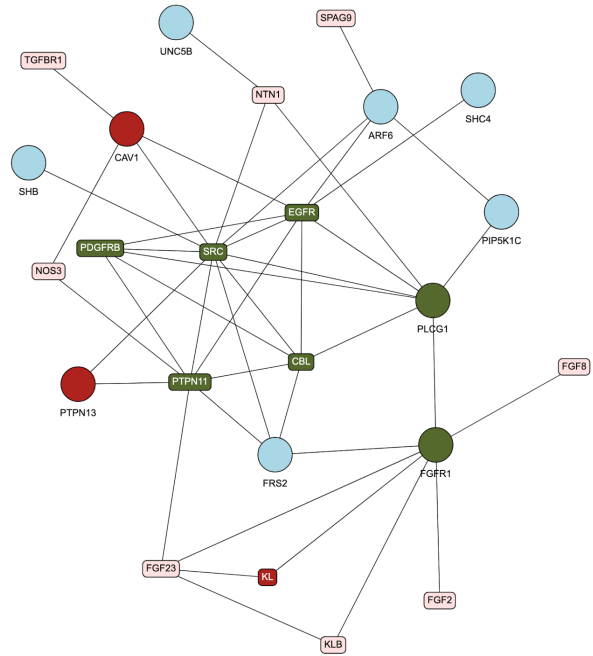

**B**

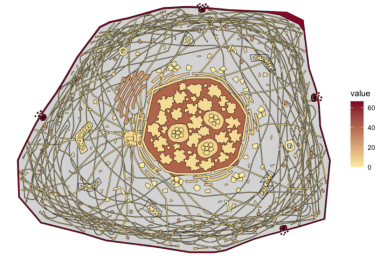

**C**

**D**

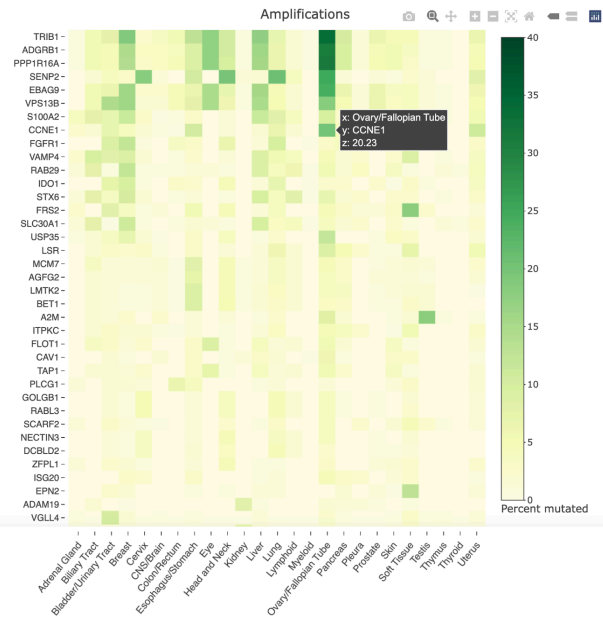

**E**

**Figure 4. Proteomics use case analysis - FGFR1 signaling network.** Specific output views from the oncoEnrichR HTML report, analyzing N = 74 proteins found through a protein proximity screen using FGFR1 as the bait (complete reports available at https://doi.org/10.5281/zenodo.7115732). **A.** Protein-protein interaction network, validating the proximal interaction partners of FGFR1. Query set members as circles, interacting non-query set proteins as squares, cancer gene roles are color-coded (tumor suppressors in green, proto-oncogenes in red, dual roles in black). **B.** Subcellular anatogram indicating main localization of hits (secretory pathway, the plasma membrane, the endosomal system and the nucleus). **C.** Cancer drugs in early or late phase development targeted against members of the query set. **D.** High-level amplifications of query set members found in tumor samples from TCGA cohorts, showcasing considerable frequency of CCNE1 amplifications in samples originating from the ovary/fallopian tube. **E.** List of genes with poorly defined or unknown functions detected in the query set, indicated through lack of gene summary descriptions or Gene Ontology (GO) terms.

*Supplementary Materials*

**Comprehensive interrogation of gene lists from genome-scale cancer screens with oncoEnrichR**

Nakken et al., 2022

## Supplementary Methods

**Dataset collections**

Here, we outline the type of knowledge databases that are integrated in oncoEnrichR, including quality control (QC) and pre-processing steps of the raw data, and also how the various datasets can be restricted/filtered (through confidence thresholds etc.) when conducting a gene set analysis.

*Target-cancer associations*

Direct associations between human genes and phenotypes/diseases were retrieved from the Open Targets Platform (OTP, https://platform.opentargets.org/downloads/data, (1). Here, associations are quantified as numerical scores in the range 0-1, with larger values indicating a stronger association between the gene and the phenotype (https://platform-docs.opentargets.org/associations#overall). Phenotypes from OTP were provided as Experimental Factor Ontology (EFO) terms. For each OTP association, we integrated the underlying data type evidence, examples being genetic associations, text mining, pathway associations, or animal models. Next, in order to focus on the subset of associations with strongest confidence in underlying evidence, we considered only associations with support from at least two data types, and with an overall association score >= 0.05. Finally, in order to limit associations to cancer phenotypes, we set up a curated mapping between EFO terms and n = 32 primary tumor types/sites, using OncoTree as a starting point (the complete mapping is available at https://github.com/sigven/oncoPhenoMap). This produced a final list of n = 33,369 associations between genes and cancer phenotype terms.

In order to establish an overall ranking of genes with respect to cancer relevance, we summarized, for each tumor type, the potential multiple OTP association scores found per gene, and computed a scaled rank between 0 and 1 (i.e. **gene-tumor type rank**). Next, we computed a global (i.e. *pancancer*) **gene-cancer rank** per gene by summarizing the **gene-tumor type rank** found across tumor types, with all genes ultimately ranked and scaled between 0 and 1 (**Supplementary Figure S1**).

*Target-drug associations*

The *Drug* dataset in JSON format was downloaded from the Open Targets Platform. This dataset contains records that link drug molecule identifiers (ChEMBL compounds) with disease indications (Experimental Factor Ontology (EFO) terms), as well as the identity of the molecular target. We further combined this dataset with the *Drug - indication* dataset from OTP, which contains information with respect to the maximum clinical trial phase of a particular drug for a given indication/disease. Finally, drugs indicated specifically for cancer phenotypes were established by intersecting the raw drug dataset from OTP with our curated mapping of phenotype terms per primary tumor type (defined above). Targeted cancer drugs were categorized as being in either early (max phase 1 or 2) or late clinical development (max phase 3 or 4). To indicate other potential prospects for drug targeting, we also included target tractability data from OTP, indicating whether proteins are generally amenable for targeting by antibodies or small molecules (https://platform-docs.opentargets.org/target/tractability).

*Tumor suppressor, proto-oncogene, and cancer driver gene annotation*

In order to annotate genes with roles as proto-oncogenes, tumor suppressors, or cancer drivers, we combined evidence from three sources; curated annotations from Cancer Gene Census (CGC v96) and Network of Cancer Genes (NCG 7.0), and literature-mining data from CancerMine (2–4). From CancerMine (v48), we downloaded all PubMed identifiers (PMIDs) for which there were sentence level evidence (probability > 0.8) for a role of a gene as tumor suppressor, proto-oncogene, or cancer driver. Next, we set up a scheme in which annotations either required support from multiple PMIDs in CancerMine, or through existing curated annotations in NCG or CGC (yet with potentially fewer supporting PMIDs in CancerMine). Specifically, the classification of proto-oncogenes/tumor suppressors was done according to the following scheme:

- Existing annotation as proto-oncogene/tumor suppressor in CGC *or* NCG, *or* evidence from >=15 distinct PMIDs (CancerMine) that suggested an oncogenic/tumor suppressor role for a given gene
- Status as proto-oncogene was ignored if a given gene had three times as much (literature evidence/PMIDs) support for a role as a tumor suppressor gene (and vice versa for tumor suppressors)
- Proto-oncogenes/tumor suppressor candidates that were found in a curated list of false positive cancer driver predictions were excluded (5).

The above strategy resulted in a total of n = 401 proto-oncogenes, n = 359 tumor suppressor genes, and with n = 83 of them classified with dual roles.

For cancer driver gene classification, we considered the union of predictions from IntOGen (v2022-02-01) and NCG 7.0 (3,6). For NCG, we only considered drivers annotated as "canonical cancer genes". Similar to the approach employed for proto-oncogenes and tumor suppressor genes, we ignored candidates that were found in a curated list of false positive driver predictions (5). Based on the outlined approach, a total of n = 722 genes were classified as cancer drivers. For each cancer driver prediction, we also appended available literature evidence from CancerMine.

*Cancer hallmark evidence*

The hallmarks of cancer comprise six biological capabilities acquired during the multistep development of human tumors (7). Cancer hallmark annotations per gene were collected from the Open Targets Platform (1), as found by parsing raw JSON files with target data (https://platform.opentargets.org/downloads/data)

*Targets of unknown function*

A dedicated section in oncoEnrichR highlights candidate targets for which the gene function is poorly characterized or unknown. We considered the following properties as evidence for a poorly characterized gene; i) missing gene summary in NCBI Gene/UniProt, or a gene name designated as "uncharacterized" or "open reading frame" , ii) limited number (<= 1) of annotated, manually reviewed (non-IEA) gene ontology terms, coming from either the molecular function (MF) or biological process (BP) sub-ontologies. A total of n = 3,122 protein-coding human genes were assigned an unknown/poorly defined function according to this scheme.

*Copy number amplifications/homozygous deletions*

Somatic copy number amplifications and homozygous deletions, coming from Affymetrix SNP 6.0 (SNP6) array data, were retrieved from The Cancer Genome Atlas (TCGA, Genomics Data Commons, release 34), using the R package TCGAbiolinks (8). We considered the thresholded copy-number status of protein-coding genes as assessed by the GISTIC2 pipeline (function *getGistic()* in TCGAbiolinks), limited to high-level amplifications (GISTIC2 CNA value of 2) and homozygous deletions (GISTIC2 CNA value equal to -2). All CNA events were organized on a per-tumor type basis.

*Recurrent SNVs/InDels*

Recurrent (occurring in more than one tumor sample) somatic point mutations and indels were collected from TCGA, using TCGABiolinks (Genomic Data Commons release 34). All mutations and indels were further annotated with known mutation hotspots (cancerhotspots.org (9), protein domains from Pfam (10), and their predicted status as loss-of-function variants, using the LOFTEE algorithm (11). Only protein-altering and canonical splice site variants were included, and variants were organized on a per-tumor type basis, allowing for variant interrogation in specific tumor types. In total, n = 240,985 recurrent, protein-coding variants from TCGA were included.

*Co-expression (RNA-seq)*

Co-expression correlation coefficients between protein-coding genes in tumors were calculated using RNAseq data available from TCGA cohorts. Raw expression values (TPM) were downloaded using TCGABiolinks (Genomic Data Commons release 34), and log-transformed (log2(TPM + 1)). Gene-vs-gene expression correlation coefficients (Spearman's rank) were calculated for each TCGA cohort. We focused specifically on co-expressed genes with a potential cancer relevance, by limiting pairs of co-expressed genes to those that included either a tumor suppressor, an oncogene, or a predicted cancer driver (i.e. using the annotations outlined in section *Tumor suppressor/oncogene/cancer driver annotation* above). Only pairs with strong correlation coefficients ($r >= 0.7$ or $r <= -0.7$) and an associated p-value < 1.0e-6 were included (in total n = 5,856,802 pairs).

*Molecular signatures / Gene Ontology (GO) / pathway annotations*

We collected pathway annotations and molecular gene signatures from multiple databases. Gene signature collections found in the Molecular Signatures Database (MSigDB (v2022.1.Hs)) were downloaded, which included multiple cancer-relevant signature collections, such as oncogenic signature gene sets, and computationally-derived cancer modules (12,13). MSigDB was also used as a source for Gene Ontology annotations. Biological pathway annotations from WikiPathways (release 2022-09-10), and KEGG (release 2022-08-09), were downloaded separately from MSigDb, ensuring most up-to-date annotations (14,15). Signaling pathway annotations from NetPath (http://www.netpath.org/) were also included (16).

All pathway and signature datasets were used for gene set enrichment or over-representation analysis of the user-defined candidate geneset. Geneset over-representation analysis was conducted with clusterProfiler (17).

*Protein-protein interactions*

Protein-protein interactions in oncoEnrichR are directly retrieved from the STRING resource, utilizing their publicly available API (https://string-db.org/cgi/help?subpage=api) for network retrieval (18). In the protein-protein interactome module of oncoEnrichR, users provide a required (minimum) interaction score threshold for proteins to be included in the network, and further a selected number of proteins (between 1 and 50), so that the network can be expanded with important interacting proteins connected to the initial network formed by the input set. Importantly, the proteins/nodes in the network that are not part of the candidate set can be distinguished by means of a different shape (squares rather than circles).

Community structures in the protein-protein interaction network are identified with the *fastgreedy.community* method in the igraph R package (19). Network centrality scores (i.e. indicating "hub" proteins with respect to interaction) are calculated according to the method by Kleinberg (20).

*Cell-type and tissue-specific expression patterns*

Tissue-specific gene expression patterns were collected from The Human Protein Atlas (HPA) (21). Consensus transcript expression levels (nTPM) summarized per gene in 55 tissues based on transcriptomics data from HPA and Gene Expression Tissue (GTEx) project were downloaded (HPA release 21). Based on the expression profile across tissues, each gene was further assigned to one of five tissue specificity categories: *tissue enriched, group enriched, tissue enhanced, low tissue specificity, not detected*. The latter was accomplished using the *teGeneRetrieval* method in the R package TissueEnrich, using a fold-change threshold of four to define *enriched*, *group enriched* and *enhanced* categories (22).

RNA single cell type data was downloaded from the Human Protein Atlas (HPA release 21), comprising transcript expression levels summarized per gene in 76 cell types from 26 datasets. Similar to the approach used for tissue types, we used TissueEnrich to classify genes into five specificity categories, i.e. cell type enriched, cell type enhanced etc.

*Gene fitness effects and target priority scores*

Gene fitness effects, as estimated from genome-scale CRISPR-Cas9 drop-out screens in > 800 human cancer cell lines, were downloaded from Project Score, released june 2021 (23). For each cell line, we collected significant *fitness scores* per gene, in the form of gene-level Bayes Factor (BF) values, which is a quantitative measure of the cell viability effect elicited by CRISPR-Cas9 mediated cell inactivation. A total of n = 1,581,786 gene and cell-line specific BF values (<0) were retrieved. In oncoEnrichR, users may set an upper threshold on the BF value for fitness data retrieved per gene.

We downloaded *target priority scores* provided per gene, which scores the potential of a gene as a therapeutic target in a given cancer type (scale 0-100, n = 15,661 scores in total). This score exploits primarily the size and significance of the CRISPR-derived fitness effects, but also the prevalence of existing genomic biomarkers (24).

*Subcellular localization*

Subcellular localization annotations were retrieved from ComPPI (25). Here, each gene is annotated towards subcellular compartments at *major* and *minor* levels. There are six major compartment levels (cytosol, nucleus, extracellular, membrane, secretory-pathway, and mitochondrion), and the annotation of a gene towards a major compartment level is provided with a score/confidence from 0-1. Within each major compartment, there can exist multiple minor level annotations in the form of GO terms, and where the confidence of each minor level annotation is indicated by the number of databases (e.g. HPA, GO, LOCATE, eSLDB etc.) that support it. In oncoEnrichR, we restrict major compartment annotations to those with a score >= 0.8. Furthermore, users have the possibility to set a minimum threshold with respect to the confidence (number of underlying databases/sources, 1-6) for each target-localization annotation occurring at the detailed/minor level.

In order to visualize the tendency of subcellular localization for a given candidate list, we utilized *gganatogram* (26). This software focuses on n = 24 subcellular structures for visualization (https://github.com/jespermaag/gganatogram#cellular-structures). We manually created a mapping between GO terms provided by ComPPI towards the subcellular structures visualized with *gganatogram*.

*Regulatory interactions*

Regulatory interactions (transcription factor - target) were retrieved from the DoroThEA resource (27); a *global set* of regulatory interactions (n = 273,572), many of which are inferred from gene expression patterns in normal tissues (GTEx), and a *cancer-focused set* (n = 213,230), the latter inferred from gene expression patterns in tumors (TCGA). In oncoEnrichR, users have the possibility to set a minimum threshold with respect to the confidence (levels A/B/C/D, with A being strongest) for the TF-target interactions shown.

*Prognostic associations*

Prognostic associations were collected from two different sources, the Human Protein Atlas (HPA), and https://survival.cshl.edu/ (28,29).

From HPA, we collected significant results (p <= 0.001) from correlation analyses of mRNA expression levels of human genes in tumor tissue versus clinical outcome (survival) for ~8,000 cancer patients (underlying data from TCGA). These data allow for the interrogation of both favorable and unfavorable genes, the latter set referring to genes for which a higher relative expression in a given tumor type gives significantly lower survival for the patients. We further downloaded recently established data from (28), in which Cox proportional hazards models were generated to link the expression, copy number, methylation, or mutation status of every gene in the genome with patient outcome in 16 different cancer types profiled by the TCGA. Here, survival associations were proved as Z scores (Wald statistic), indicating whether the coefficient of a given variable (e.g. gene CNA status) is significantly different from zero.

*Protein complexes*

We collected data on human protein complexes from OmniPath, which contains records from multiple databases, primarily CORUM, ComplexPortal, Compleat, and hu.MAP2 (30–34). We removed several duplicate entries, and restricted complexes to those with more than two participating proteins (resulting in a total of n = 8,096 complexes). Since the hu.MAP2 resource is based solely on computational predictions, we gathered these entries in a separate collection, so that the user could explore these independently from the other (curated) entries. Finally, we assigned a score per complex, indicating the overall cancer relevance of the participating members, specifically the mean pancancer gene rank (as outlined above in the section *Target-cancer associations*).

*Ligand-receptor interactions*

Ligand-receptor interactions (n = 1,939) were collected from the CellChatDB resource (35). In the oncoEnrichR module for such interactions, we only show interactions where both ligand and receptor are found in the candidate set.

*Synthetic lethality predictions*

We downloaded recently published predictions for synthetic lethality interactions in human cancer cell lines (36). Here, human gene paralogs were predicted as potentially synthetic lethal based on a number of features, most prominently shared protein-protein interactions, and percent sequence identity. A total of n = 36,648 predictions were provided with a prediction score and percentile, indicating how strong the evidence is for a potential synthetic lethality interaction. In oncoEnrichR, we show synthetic lethality interactions in which both members are part of the candidate set, as well as instances where only one member of the interaction is part of the candidate set.

*Supplementary Results*

*Use case reports*

We explored the usability of the oncoEnrichR workflow for two candidate lists, one coming from a CRISPR screen investigating drug resistance in lung cancer, and another coming from a proteomics screen probing the interacting partners of FGFR1 upon stimulation with the activating ligand FGF1.

The complete HTML reports and Excel output of both candidate lists can be downloaded from the following URL: https://doi.org/10.5281/zenodo.7115732

*References*

1. Ochoa D, Hercules A, Carmona M, Suveges D, Gonzalez-Uriarte A, Malangone C, et al. Open Targets Platform: supporting systematic drug-target identification and prioritisation. Nucleic Acids Res. 2021;49:D1302–10.

2. Lever J, Zhao EY, Grewal J, Jones MR, Jones SJM. CancerMine: a literature-mined resource for drivers, oncogenes and tumor suppressors in cancer. Nat Methods. 2019;16:505–7.

3. Repana D, Nulsen J, Dressler L, Bortolomeazzi M, Venkata SK, Tourna A, et al. The Network of Cancer Genes (NCG): a comprehensive catalogue of known and candidate cancer genes from cancer sequencing screens. Genome Biol. 2019;20:1.

4. Sondka Z, Bamford S, Cole CG, Ward SA, Dunham I, Forbes SA. The COSMIC Cancer Gene Census: describing genetic dysfunction across all human cancers. Nat Rev Cancer. 2018;18:696–705.

5. Bailey MH, Tokheim C, Porta-Pardo E, Sengupta S, Bertrand D, Weerasinghe A, et al. Comprehensive Characterization of Cancer Driver Genes and Mutations. Cell. 2018;173:371–85.e18.

6. Martínez-Jiménez F, Muiños F, Sentís I, Deu-Pons J, Reyes-Salazar I, Arnedo-Pac C, et al. A compendium of mutational cancer driver genes. Nat Rev Cancer. Nature Publishing Group; 2020;20:555–72.

7. Hanahan D, Weinberg RA. Hallmarks of cancer: the next generation. Cell. 2011;144:646–74.

8. Colaprico A, Silva TC, Olsen C, Garofano L, Cava C, Garolini D, et al. TCGAbiolinks: an R/Bioconductor package for integrative analysis of TCGA data. Nucleic Acids Res. 2016;44:e71.

9. Chang MT, Asthana S, Gao SP, Lee BH, Chapman JS, Kandoth C, et al. Identifying recurrent mutations in cancer reveals widespread lineage diversity and mutational specificity. Nat Biotechnol. 2016;34:155–63.

10. Finn RD, Bateman A, Clements J, Coggill P, Eberhardt RY, Eddy SR, et al. Pfam: the protein families database. Nucleic Acids Res. 2014;42:D222–30.

11. Karczewski KJ, Francioli LC, Tiao G, Cummings BB, Alföldi J, Wang Q, et al. The mutational

constraint spectrum quantified from variation in 141,456 humans. Nature. 2020;581:434–43.

12. Subramanian A, Tamayo P, Mootha VK, Mukherjee S, Ebert BL, Gillette MA, et al. Gene set enrichment analysis: a knowledge-based approach for interpreting genome-wide expression profiles. Proc Natl Acad Sci U S A. 2005;102:15545–50.

13. Liberzon A, Subramanian A, Pinchback R, Thorvaldsdóttir H, Tamayo P, Mesirov JP. Molecular signatures database (MSigDB) 3.0. Bioinformatics. 2011;27:1739–40.

14. Kanehisa M, Goto S. KEGG: kyoto encyclopedia of genes and genomes. Nucleic Acids Res. 2000;28:27–30.

15. Slenter DN, Kutmon M, Hanspers K, Riutta A, Windsor J, Nunes N, et al. WikiPathways: a multifaceted pathway database bridging metabolomics to other omics research. Nucleic Acids Res. Oxford Academic; 2017;46:D661–7.

16. Kandasamy K, Mohan SS, Raju R, Keerthikumar S, Kumar GSS, Venugopal AK, et al. NetPath: a public resource of curated signal transduction pathways. Genome Biol. 2010;11:R3.

17. Yu G, Wang L-G, Han Y, He Q-Y. clusterProfiler: an R package for comparing biological themes among gene clusters. OMICS. 2012;16:284–7.

18. Von Mering C, Jensen LJ, Snel B, Hooper SD, Krupp M, Foglierini M, et al. STRING: known and predicted protein--protein associations, integrated and transferred across organisms. Nucleic Acids Res. Oxford University Press; 2005;33:D433–7.

19. Clauset A, Newman MEJ, Moore C. Finding community structure in very large networks. Phys Rev E Stat Nonlin Soft Matter Phys. 2004;70:066111.

20. Kleinberg JM. Authoritative sources in a hyperlinked environment. J ACM. New York, NY, USA: Association for Computing Machinery; 1999;46:604–32.

21. Uhlén M, Fagerberg L, Hallström BM, Lindskog C, Oksvold P, Mardinoglu A, et al. Proteomics. Tissue-based map of the human proteome. Science. 2015;347:1260419.

22. Jain A, Tuteja G. TissueEnrich: Tissue-specific gene enrichment analysis. Bioinformatics. 2019;35:1966–7.

23. Behan FM, Iorio F, Picco G, Gonçalves E, Beaver CM, Migliardi G, et al. Prioritization of cancer therapeutic targets using CRISPR-Cas9 screens. Nature. 2019;568:511–6.

24. Dwane L, Behan FM, Gonçalves E, Lightfoot H, Yang W, van der Meer D, et al. Project Score database: a resource for investigating cancer cell dependencies and prioritizing therapeutic targets. Nucleic Acids Res. Oxford Academic; 2020;49:D1365–72.

25. Veres DV, Gyurkó DM, Thaler B, Szalay KZ, Fazekas D, Korcsmáros T, et al. ComPPI: a cellular compartment-specific database for protein-protein interaction network analysis. Nucleic Acids Res. 2015;43:D485–93.

26. Maag JLV. gganatogram: An R package for modular visualisation of anatograms and tissues based on ggplot2. F1000Res. 2018;7:1576.

27. Garcia-Alonso L, Holland CH, Ibrahim MM, Turei D, Saez-Rodriguez J. Benchmark and integration of resources for the estimation of human transcription factor activities. Genome

Res. 2019;29:1363–75.

28. Smith JC, Sheltzer JM. Genome-wide identification and analysis of prognostic features in human cancers. Cell Rep. 2022;38:110569.

29. Uhlen M, Zhang C, Lee S, Sjöstedt E, Fagerberg L, Bidkhori G, et al. A pathology atlas of the human cancer transcriptome. Science [Internet]. 2017;357. Available from: http://dx.doi.org/10.1126/science.aan2507

30. Giurgiu M, Reinhard J, Brauner B, Dunger-Kaltenbach I, Fobo G, Frishman G, et al. CORUM: the comprehensive resource of mammalian protein complexes—2019. Nucleic Acids Res. Narnia; 2019;47:D559–63.

31. Meldal BHM, Bye-A-Jee H, Gajdoš L, Hammerová Z, Horáčková A, Melicher F, et al. Complex Portal 2018: extended content and enhanced visualization tools for macromolecular complexes. Nucleic Acids Res. Oxford Academic; 2018;47:D550–8.

32. Vinayagam A, Hu Y, Kulkarni M, Roesel C, Sopko R, Mohr SE, et al. Protein complex-based analysis framework for high-throughput data sets. Sci Signal. 2013;6:rs5.

33. Drew K, Wallingford JB, Marcotte EM. hu.MAP 2.0: integration of over 15,000 proteomic experiments builds a global compendium of human multiprotein assemblies. Mol Syst Biol. 2021;17:e10016.

34. Türei D, Korcsmáros T, Saez-Rodriguez J. OmniPath: guidelines and gateway for literature-curated signaling pathway resources. Nat Methods. 2016;13:966–7.

35. Jin S, Guerrero-Juarez CF, Zhang L, Chang I, Ramos R, Kuan C-H, et al. Inference and analysis of cell-cell communication using CellChat. Nat Commun. 2021;12:1088.

36. De Kegel B, Quinn N, Thompson NA, Adams DJ, Ryan CJ. Comprehensive prediction of robust synthetic lethality between paralog pairs in cancer cell lines. Cell Syst. 2021;12:1144–59.e6.

*Supplementary Figures*

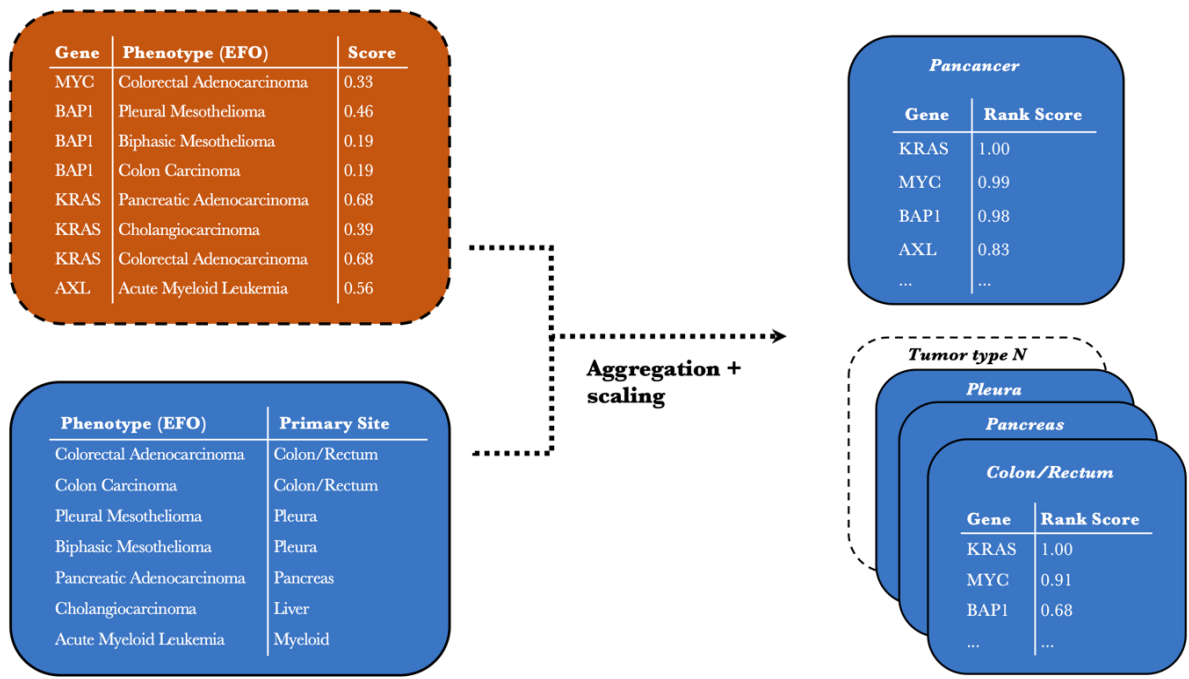

**Supplementary Figure S1. Generation of gene-cancer association ranks.** By combining gene-phenotype term association scores from the Open Targets Platform (upper left) with a curated mapping of phenotype terms toward primary tumor sites (https://github.com/sigven/oncoPhenoMap), we established a global/pan-cancer rank of genes with respect to cancer relevance, and also a tumor-type specific ranking, the latter reflecting the strength of association between a gene and a particular tumor type.

**Supplementary Tables**

**Table 1**: Overview of Scientific Questions addressed by oncoEnrichR

https://docs.google.com/spreadsheets/d/1SBQ0Yn5pYB3XzxwJIsayTvwT71RHPxC9/

**Table 2**: Detailed overview of annotation resources used in oncoEnrichR, including versions, publications, update frequencies, and quality control procedures

https://docs.google.com/spreadsheets/d/1SBQ0Yn5pYB3XzxwJIsayTvwT71RHPxC9/